\documentclass[submission, Phys]{SciPost}

\hypersetup{
    colorlinks,
    linkcolor={red!50!black},
    citecolor={blue!50!black},
    urlcolor={blue!80!black}
}

\usepackage[bitstream-charter]{mathdesign}
\DeclareSymbolFont{usualmathcal}{OMS}{cmsy}{m}{n}
\DeclareSymbolFontAlphabet{\mathcal}{usualmathcal}

\begin{document}

\begin{center}{\Large \textbf{
A universal training scheme and the resulting universality for machine learning phases\\
}}\end{center}

\begin{center}
Y.-H. Yseng\textsuperscript{1},
F.-J. Jiang\textsuperscript{2$\star$} and
C.-Y. Huang\textsuperscript{3$\dagger$}
\end{center}

\begin{center}
{\bf 1} Department of Physics, National Taiwan Normal University,
88, Sec.4, Ting-Chou Rd., Taipei 116, Taiwan
\\
{\bf 2} Department of Physics, National Taiwan Normal University,
88, Sec.4, Ting-Chou Rd., Taipei 116, Taiwan
\\
{\bf 3} Department of Applied Physics,Tunghai University, No.~1727, Sec.~4, Taiwan Boulevard, Xitun District, Taichung 40704, Taiwan
\\
${}^\star$ {\small \sf fjjiang@ntnu.edu.tw}
${}^\dagger$ {\small \sf cyhuang@thu.edu.tw}
\end{center}

\begin{center}
\today
\end{center}


\section*{Abstract}
{\bf
An autoencoder (AE) and a generative adversarial networks (GANs) are trained
only once on a one-dimensional (1D) lattice of 200 sites.
Moreover, the AE contains only one hidden layer consisting of two
neurons and both the generator and the discriminator of the GANs are made up of two
neurons as well. The training set employed to train both the considered
unsupervised neural networks (NN)
is composed of two artificial configurations. Remarkably, despite their
simple architectures, both the built AE and GANs have precisely
determined the critical points of several models, including the three-dimensional (3D)
classical $O(3)$ model, the two-dimensional (2D) generalized classical XY
model, the 2D two-state Potts model, and the 1D Bose-Hubbard model.
The results presented here as well as that shown in {\it Eur. Phys. J. Plus {\bf 136}, 1116 (2021)}
suggest that when phase transitions are considered, an elegant universal
neural network that is extremely efficient and is applicable to broad physical systems can be constructed with ease. In particular,
since a NN trained with two configurations can be applied to many models, it is likely that when machine learning is concerned, the majority of phase transitions belong to a class having two elements, i.e. the Ising class.
}

\vspace{10pt}
\noindent\rule{\textwidth}{1pt}
\tableofcontents\thispagestyle{fancy}
\noindent\rule{\textwidth}{1pt}
\vspace{10pt}

\section{Introduction}

Machine learning (ML) techniques have gained dramatic attention recently in various fields of
physics. Such examples include astronomy, particle physics, computational materials, and condensed matter physics \cite{Sny12,Li15,Baldi:2014pta,Baldi:2016fql,Hoyle:2015yha,Tor16,Wan16,Mott:2017xdb,Bro16,Car16,Tubiana:2016zpw,Kol17,Nie16,Tro16,Nag17,Den17,Zha17,Hu17,Zha17.1,Tan17,Bea18,Chn18,Pang:2016vdc,Lu18,Kim18,But18,Bar18,Sha18,Li18,Meh19,Car19,Zha19,Gre19,Don19,Kas19,Ale20,Larkoski:2017jix,Han:2019wue,Tan20.1,Aad:2020cws,Morgan:2020wvf,Tan20.2,Lidiak:2020vgk,Nicoli:2020njz,Tan21,Tse22}.
While these applications are still in the exploratory stage, great improvement over the
traditional approaches may be expected in the near future. Among the commonly used ML
methods, neural networks (NNs), both supervised and unsupervised ones, are adopted in
distinguishing various states of matters. They are considered in reproducing certain
classical statistical distributions as well. Due to the many successful achievements of utilizing ML and NNs methods for physical problems,
it is anticipated optimistically that promising breakthroughs in certain research fields of physics may be right around the corner. 

One of the applications of NNs in physical problems is to detect different phases of matters. Indeed,
for many systems and models, both supervised and unsupervised NNs
have explored the associated phase diagrams with success to certain extents.
Between the supervised and the unsupervised NNs, unsupervised ones are typically favored. This is because prior information is needed
before one can apply the techniques of supervised NNs.
As a result, the majority of NN studies for physical systems typically use unsupervised approaches.

While unsupervised NNs have dominated the applications of NNs in physical problems, supervised
NNs have certain advantage over the unsupervised ones. For instance, supervised NNs are easier
to implemented and often lead to better outcomes. Based on these facts, it is benefited
to continue examining the potential applications of supervised NNs in reality.  

Typically, applying a NN to study the properties of many-body systems has three stages, namely the training, the validation,
and the testing (prediction) stages. Among these three stages, the training is the most time consuming one. In particular,
with the conventional training strategy usually employed in the literature, a new training is required whenever
a new model or a different system size is considered. It should be pointed out that the configurations employed
for the training are based on physical quantities such as the spins or the correlation functions. As a result, certain
amount of computing time is needed to generate the relevant physical data in order to construct the required
configurations for the NN training. These mentioned technical difficulties usually prevent the NNs
from being used in real problems practically.

We would like to emaphsize the fact that in the literature the NNs used to explore physical problems usually have very complicated
architectures such as the deep learning convolutional neural networks (CNN). Such a strategy of building the NNs
would lead to a lot of trainable (tunable) NN parameters. As a result, huge amount of computing time is required
in order to obtained a working NN that has a dedicated architecture.

In Ref.~\cite{Li18}, the two-dimensional (2D) $q$-state ferromagnetic Potts models on the square lattice are studied.
Unlike the conventional training procedure which uses real Potts configurations generated by some
numerical methods as the training set, the theoretical ground state configurations are employed instead. 
With this training strategy which costs much less computing time, accurate determinations of the critical points are obtained.
Later in Refs.~\cite{Tan21}, two artificially made configurations on a one-dimensional (1D) lattice of 120 sites are used as the training sets, and
the resulting (only one) supervised NN (which contains one hidden layer) has successfully determined the critical points of the
three-dimensional (3D) classical $O(3)$ model, the 2D classical $XY$ (generalized classical $XY$ model),
a 3D dimerized spin-1/2 Heisenberg model, and the 3D 5-state ferromagnetic Potts model \cite{Tan21,Tse22}. It is remarkable that the (only one)
simple supervised NN constructed in Ref.~\cite{Tan21} is capable of studying the phase transitions of various 3D and 2D models.
In particular, prior information is not necessary to carry out the investigations presented in Refs.~\cite{Tan21,Tse22}.

Although the fact that no information of the considered system is needed to utilize the supervised NN built in Ref.~\cite{Tan21} makes the necessarity of using unsupervised NNs no long warranted, 
it is still interesting to examine whether the simple and elegant
training approach in Ref.~\cite{Tan21} can be adopted to train unsupervised NNs. Because of this motivation, here
we train an autoencoder (AE) and a generative adversarial networks (GANs) using two artificially made
configurations on a 1D lattice of 200 sites. In particular, besides the input and the output layers,
the AE built here contains only one hidden layer consisting of 2 neutrons.
Moreover, each of the generator and discriminator of the constructed GANs
has only 2 neurons as well.

Remarkably, despite their simple architectures, the AE and the GANs obtained here with the elegant training strategy
successfully estimate the critical points of various models, including the 3D classical $O(3)$ model,
the 2D 2-state Potts model, the 2D classical generalized $XY$ model, as well as
the 1D Bose-Hubbard model. We would like to emphasize the fact that both of these two unsupervised NNs have only been trained once, and
no new training is conducted whenever new models or different system sizes are considered.
It is amazing that two NNs, each of which in principle has only two neurons and is trained on a 1D lattice, can be adopted to calculate the critical points of many 3D, 2D and 1D models
with high precision. In particular, no information of these stuided models
are needed for these two built unsupervised NNs to detect the associated
phase transitions. Some benchmark investigations are carried out as well 
to demonstrate the efficiency of the NN(s) constructed here.

We would like to point out that with minor modification for the testing sets,
it is likely the NNs obtained here and in
Refs.~\cite{Tan21,Tse22} can be directly applied to study other models such as the Fermi-Hubbard model as well as the
Su-Schrieffer–Heeger (SSH) model. 

Finally, it is worthy to mention that the results shown here and in Ref.~\cite{Tse22} indicate that a NN trained with only two (artificial)
configurations can be applied to many physical models that vary from each other
significantly. This implies that when machine learning is concerned, it is 
likely that the majority of phase transitions belong 
to a class having two elements, i.e. the Ising class.

The rest of the paper is organized as follows. After the introduction, the
studied models are brief described in Sec. II. In Sec. III the employed
unsupervised AE and GANs, the training strategy, and the built configurations for
NN predictions are outlined. Benchmark calculations are demonstrated in Sec. IV
and the obtained NN outcomes are presented in Sec. V. Sec. VI contains our discussions and conclusions.

\section{The considered models}

The Hamiltonians of the models considered here have the following expressions
\begin{enumerate}
	\item{3D classical $O(3)$ model \cite{Hol92,Cam02}:
		\begin{equation}
		\beta H_{O(3)} = -\beta \sum_{\left< ij\right>} \vec{s}_i\cdot\vec{s}_j,
		\label{eqn1}
		\end{equation}
		where $\beta$ is the inverse temperature and $\left< ij \right>$ stands for
		the nearest neighbor sites $i$ and $j$. In addition, 
		$\vec{s}_i$ appearing above is a unit vector belonging to a 3D sphere $S^3$ and is located at
		site $i$.
	}
	\item{2D classical generalized $XY$ model \cite{Can14,Can16}:
		\begin{equation}
		H_{\text{GXY}} = \sum_{\left< ij\right>} -\Delta\cos\left(\theta_i - \theta_j\right) - \left(1-\Delta\right)\cos\left(q\theta_i-q\theta_j\right),
		\label{eqn2}
		\end{equation}
		where the summation is over the nearest neighbors $i$ and $j$, $q$ is some
		(positive) integer, and $0\le \theta_i \le 2\pi$. We will consider the case of $\Delta = 0.25$
		and $q=3$ in this study.
	}
	\item{2D 2-state Potts model \cite{Wu82}:
		\begin{equation}
		\beta H_{\text{Potts}} = -\beta \sum_{\left< ij\right>} \delta_{\sigma_i,\sigma_j}.
		\label{eqn3}
		\end{equation}
		Here $\delta$ refers to the Kronecker function and the Potts variable $\sigma_i$
		at each site $i$ takes an integer value from $\{1,2\}$. 
	}
	\item{1D Bose-Hubbard model:
		Using the creation and annihilation operators $\hat{a}_i^{\dagger}$ and $\hat{a}_i$, the Hamiltonian of the 1D Bose-Hubbard model is given by \cite{Eji11,Eji12}
		\begin{eqnarray}
		H_{\text{BH}} = -t\sum_{i=1}^{L}\left(\hat{a}_i^{\dagger} \hat{a}_{i+1} + \hat{a}_{i+1}^{\dagger}\hat{a}_{i} \right) + \frac{U}{2}\sum_{i=1}^{L}\hat{n}_i\left(\hat{n}_i-1\right)-\mu\sum_{i=1}^{L}\hat{n}_i,
		\end{eqnarray}
		where $L$ is the number of sites, $t$ is the tunneling strength, $U > 0$ is
		the on-site repulsive interaction strength, $\mu$ is the chemical potential, and finally $\hat{n}_i = \hat{a}_i^{\dagger} \hat{a}_{i}$ is the
		particle number operator at site $i$.
	}
\end{enumerate}

\section{The constructed unsupervised NNs}

Typically, a NN has many tunable parameters. In this investigation we use the default values for these parameters unless specified.

\subsection{Autoencoder}

\begin{figure}
	\begin{center}
		\includegraphics[width=0.45\textwidth]{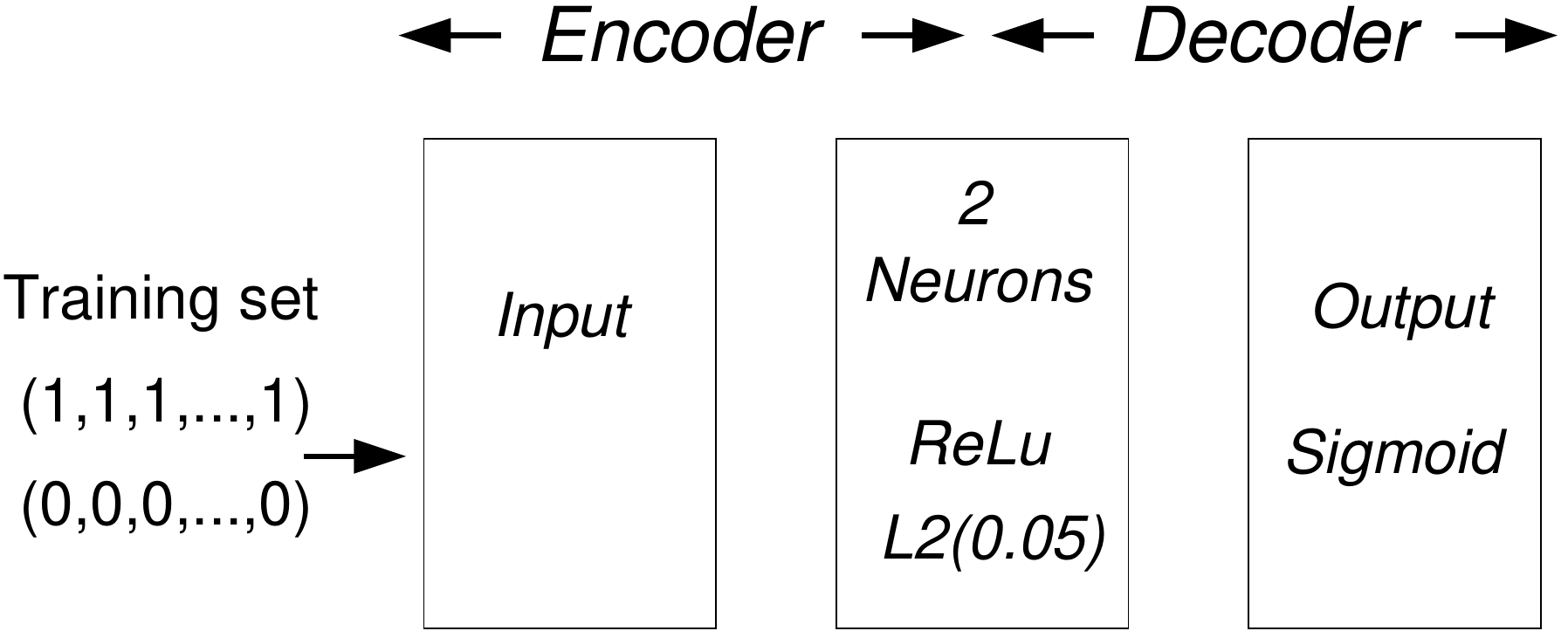}
	\end{center}\vskip-0.7cm
	\caption{The AE employed in this study. The used AE consists of one input layer, one hidden layer of
		two neurons, and one output layer. The considered activation functions (and regularizer) are shown inside the boxes.}
	\label{AE_2neurons}
\end{figure}

Using keras and tensorflow \cite{keras}, the autoencoder employed here consists of one input layer, one hidden layer
having two neurons, and one output layer. In particular, the hidden and the output layers are
activated by the ReLU and the sigmoid functions, respectively. The definitions of these two activation functions are as
follows
\begin{eqnarray}
\text{ReLU}(x) &=& \text{max}(0,x),\\
\text{sigmoid}(x) &=& \frac{1}{1+e^{x}}.   
\end{eqnarray}
The algorithm used for the training is minibatch, and one-hot encoding as well as $L_2$ regularizations 
are also applied. We use 1000 epochs and the batch size is set to 30.
Finally, the used loss function and optimizer are the crossentropy $C$ and the adam, respectively. Here
the definition of crossentropy $C$ is given by
\begin{equation}
C = -\frac{1}{n} \sum_{x}y_x \ln b_x + (1-y_x)\ln (1-b_x),
\end{equation}
where $n$ is the total number of objects in the training set, 
$b$ are the outcomes obtained after applying all the constructed layers. 
In addition, $x$ and $y$ are training inputs and
the corresponding designed outputs, respectively. 
Figure \ref{AE_2neurons} is a cartoon representation of the built AE.

For the training of the constructed AE, 200 copies of two artificially made configurations on a 1D lattice of 200
sites are used as the training set.
Specifically, every site of 200 1D configurations takes the value of 1, and each of the other 200 configurations has 0 as the values for all
of its elements.
As we will demonstrate shortly, such a simple training set up can lead to a valid NN for learning various phases of all the considered models. 
For the reader who are interested in these machine learning terminologies associated with AE (and GANs which will be introduced later)
are referred to Refs. \cite{Ras19,Ger19}.

The magnitude $R$ of the outputs is employed as the quantity for studying
the targeted phase transitions. One expects that when $R$ are considered as functions of temperature $T$ (or $\beta$), the associated outcomes
will reveal the information relevant to $T_c$ (or $\beta_c$). Indeed, for the configurations obtained at low temperatures, they are of high
similarity to the training sets. As a result, the associated outputs $R$ have large magnitude. As the temperature $T$ rises, the magnitude
$R$ of the outputs will decrease sharply at the critical point(s). As we will show later, this phenomenon is exactly what we have observed.

\subsection{Generative adversarial networks}

\begin{figure}
	\begin{center}
		\includegraphics[width=0.7\textwidth]{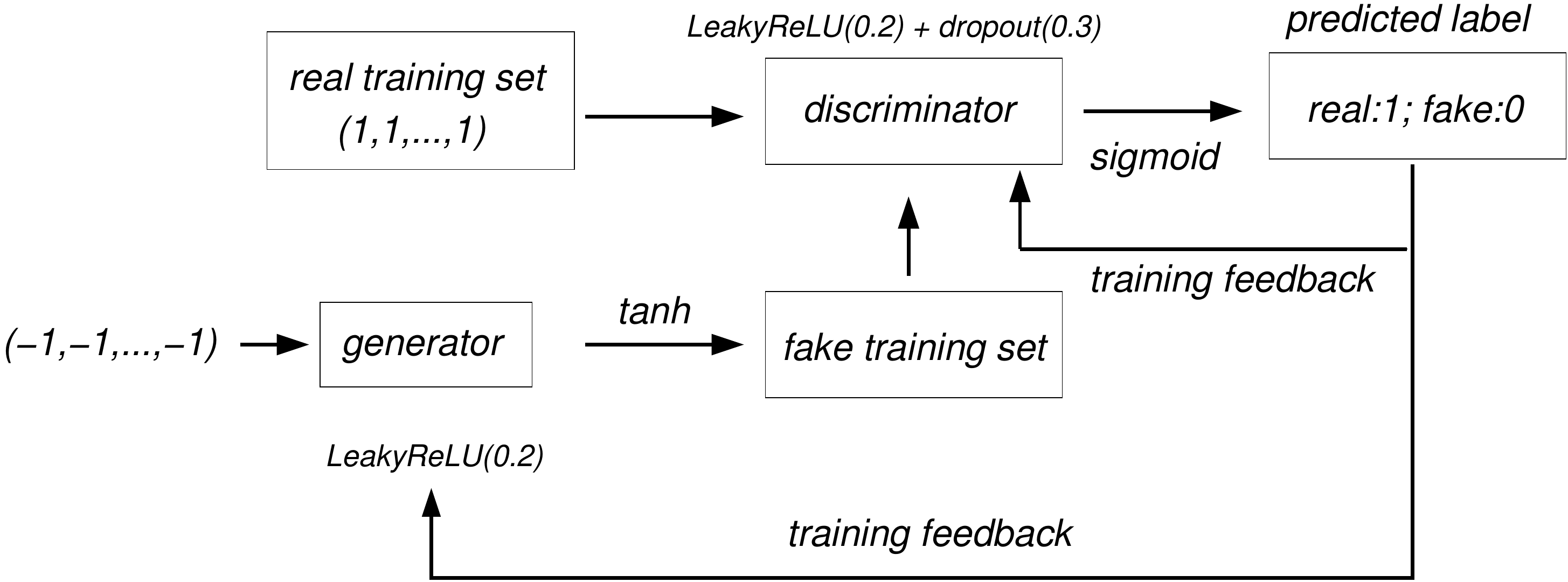}
	\end{center}\vskip-0.7cm
	\caption{The GANs employed in this study.}
	\label{GANs_2neurons}
\end{figure}

The generative adversarial networks (GANs) considered here consists of a generator with 2 neurons and a discriminator with
2 neurons as well. Each of the 640 configurations for the generator consists of 200 sites and every site has -1 as its element.
In addition, the training set is made up of 640 indentical copies of 200 sites with all the elements being 1.
Adam and crossentropy are the optimizer and the loss function employed,
respectively.
2000 epochs is carried out and we use 128 as the batch size.
The activation functions used are LeakyReLU, tanh, and sigmoid \cite{Ras19,Ger19}. The new
activation functions LeakyReLU and tanh are defined as
\begin{eqnarray}
\text{LeakyReLU}(x) &=&\left\{
\begin{array}{ll}
\alpha x  & \mbox{if } x < 0 \\
x & \mbox{if } x \ge 0
\end{array}
\right.\\
\text{tanh}(x) &=& \frac{e^{x}-e^{-x}}{e^{x}+e^{-x}},
\end{eqnarray}
where $\alpha$ is some constant.
Finally, the algorithm used for the training is again minibatch and we also apply
dropout at the discriminator. The predicted label is either 1 (true) or 0 (fake).
Figure~\ref{GANs_2neurons} is a cartoon representation of the GANs considered in this study.

Due to the data used for the fake and real training sets as well as the employed predicted label(s),
the standard deviations (STD) of the outputs will be employed to explore
the targeted phase transitions. One expects that the behavior of STD with respect to $T$ (or $\beta$) will disclose
the information relevant to $T_c$ (or $\beta_c$). In particular, for configurations obtained at low temperatures, the related GANs outputs
are either 1 or 0 with almost equal probabilities. Hence the associated outputs have large STD. As temperature $T$ rises. The magnitude of
STD will diminish dramatically at $T_c$.

\section{The benchmark calculations}

\begin{figure}
	\begin{center}
		\includegraphics[width=0.9\textwidth]{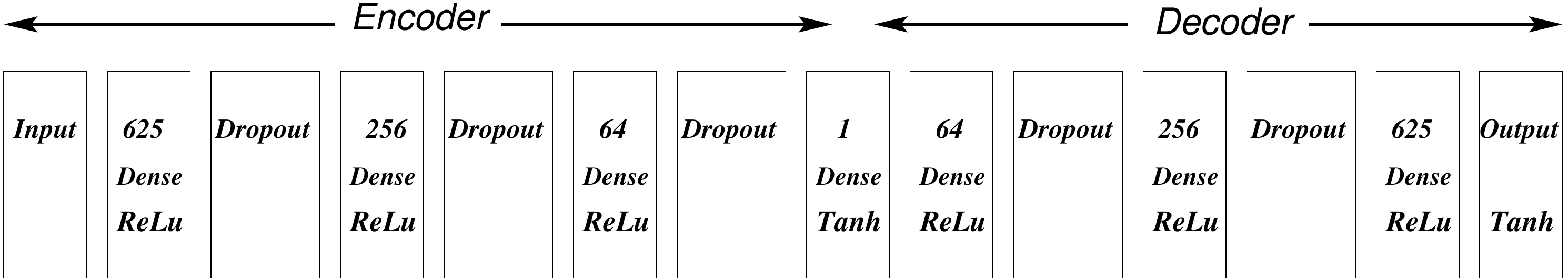}
	\end{center}\vskip-0.7cm
	\caption{The deep learning AE employed in Ref.~\cite{Ale20}. The considered activation functions and the number of neurons used are shown inside
		the boxes.}
	\label{DLAE}
\end{figure}

To demonstrate the efficiency of the AE constructed here, we have carried out certain benchmark calculations.
In particular, the deep learning autoencoder (DLAE) built in Ref.~\cite{Ale20} is employed here for the benchmark
calculations (see figure \ref{DLAE}).
The model considered is the two-state ferromagnetic Potts model on the square lattice.
In addition, the epochs and batch size used are 2000 and 30, respectively. 
Finally,
the benchmark calculations are conducted on a server with two opetron 6344 and 96G memory.

The time required to train the AE with one hidden layer having two neurons is about 110 seconds, 
while the training using 30200 real (and full) Potts configurations obtained on 128 by 128 lattices as the training set for the DLAE shown in
figure \ref{DLAE} takes about 302114 seconds to finish (It is anticipated that it will take longer to conduct all the DLAE trainings if
various system sizes are considered). Here the number of configurations used for the conventional training is about the same as that used
in Ref.~\cite{Ale20}.
Of course, the exact numbers of the benchmark calculations may depend on some
factors such as whether CPU or GUP is considered for the execution of the calculations. 
Still, the results provide certain useful information regarding the performance
of both the deep learning AE of Ref.~\cite{Ale20} and the shallow AE used in this study.
In any case, based on the benchmark outcomes, it is beyond doubt that the
unconventional training strategy adopted here is more efficient than
that typically used in the literature. Indeed, as we will demonstrate shortly,
the precision of the determined critical points using the unsupervised NN built here
is impressive as well.

\section{The numerical results}

To examine the validity as well as the efficiency of the constructed AE and GANs, certain configurations of the considered
models are required. For the classical spin models these needed configurations are generated using Monte Carlo simulations with the Wolf cluster algorithm \cite{Wol89}.
For the 1D Bose-Hubbard model, the worm algorithm of ALPS is considered \cite{alps}.

\subsection{Constructing the required configurations for predictions}

Since the configurations used for the NN predictions should have the same
dimensionality as those of the training sets, the needed
configurations for the NN predictions using the AE and GANs 
constructed here should consist of 1D lattices of 200 sites.

For the $O(3)$ model, the configurations built for the predictions are based on
the spin variable $\phi$. Specifically, for a given $O(3)$ spin configuration
obtained from the Monte Carlo simulations, 200 spins are chosen
randomly and uniformly. In addition, the $\phi$ mod $\pi$ of these 200 picked
spins are used to construct the 1D configuration for the NN predictions associated with AE.

The same procedure is applied for the 2D generalized $XY$ model. In particular,
the $\theta$ mod $\pi$ of 200 spins, which are randomly picked from an original full spin
configurations, are used as the elements of a 1D configuration for the predictions related to AE.

For every produced full Potts configuration of the two-state Potts model,
200 Potts variables $\sigma$ are chosen randomly and the associated results of $\sigma-1$ are employed to built the 1D
configuration needed for the AE predictions.

Finally, for the 1D Bose-Hubbard model, the configurations used for the NN predictions are
based on the local density.

Similar strategy is utilized to build the required configurations for the prediction related to GANs as well.

\subsection{3D classical $O(3)$ model}

\subsubsection{AE results}

\begin{figure}
	\begin{center}
		\includegraphics[width=0.5\textwidth]{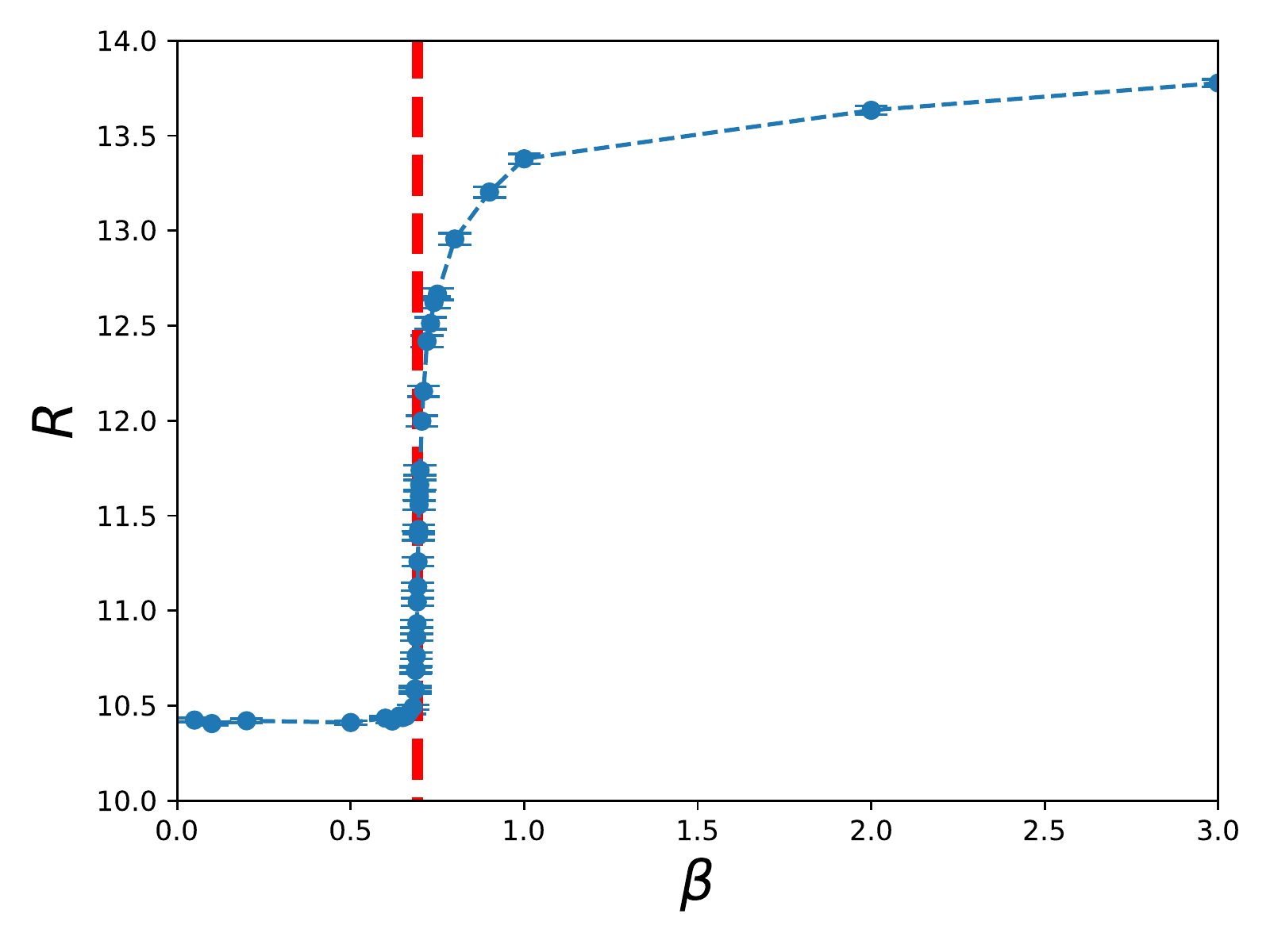}
	\end{center}\vskip-0.7cm
	\caption{The magnitude $R$ of the AE outputs as a function of the inverse temperature $\beta$ for the
		3D classical $O(3)$ model. The system size is $L$=48 and the vertical dashed line is the expected $\beta_c$.
		The value of $R$ begins to rise significantly at $\beta_c$.}
	\label{AE_O3}
\end{figure}

The magnitude $R$ of the AE outputs (In this study all the AE outputs are two hundrend components vectors) as a function of the inverse temperature $\beta$ for the
3D classical $O(3)$ model is depicted in figure \ref{AE_O3}. The system size is $L$=48 and the vertical
dashed line in that figure is the expected critical inverse temperature $\beta_c$. The outcomes given in the figure imply
that the value of $R$ begins to rise significantly at $\beta_c$. In other words, how $R$ behaves with
respect to $\beta$ reveals relevant information regarding the critical point. Particularly, $\beta_c$ can be estimated to be
the location in the associated parameter space that dramatical change of $R$ takes place.

\subsubsection{GANs results}

\begin{figure}
	\begin{center}
		\includegraphics[width=0.5\textwidth]{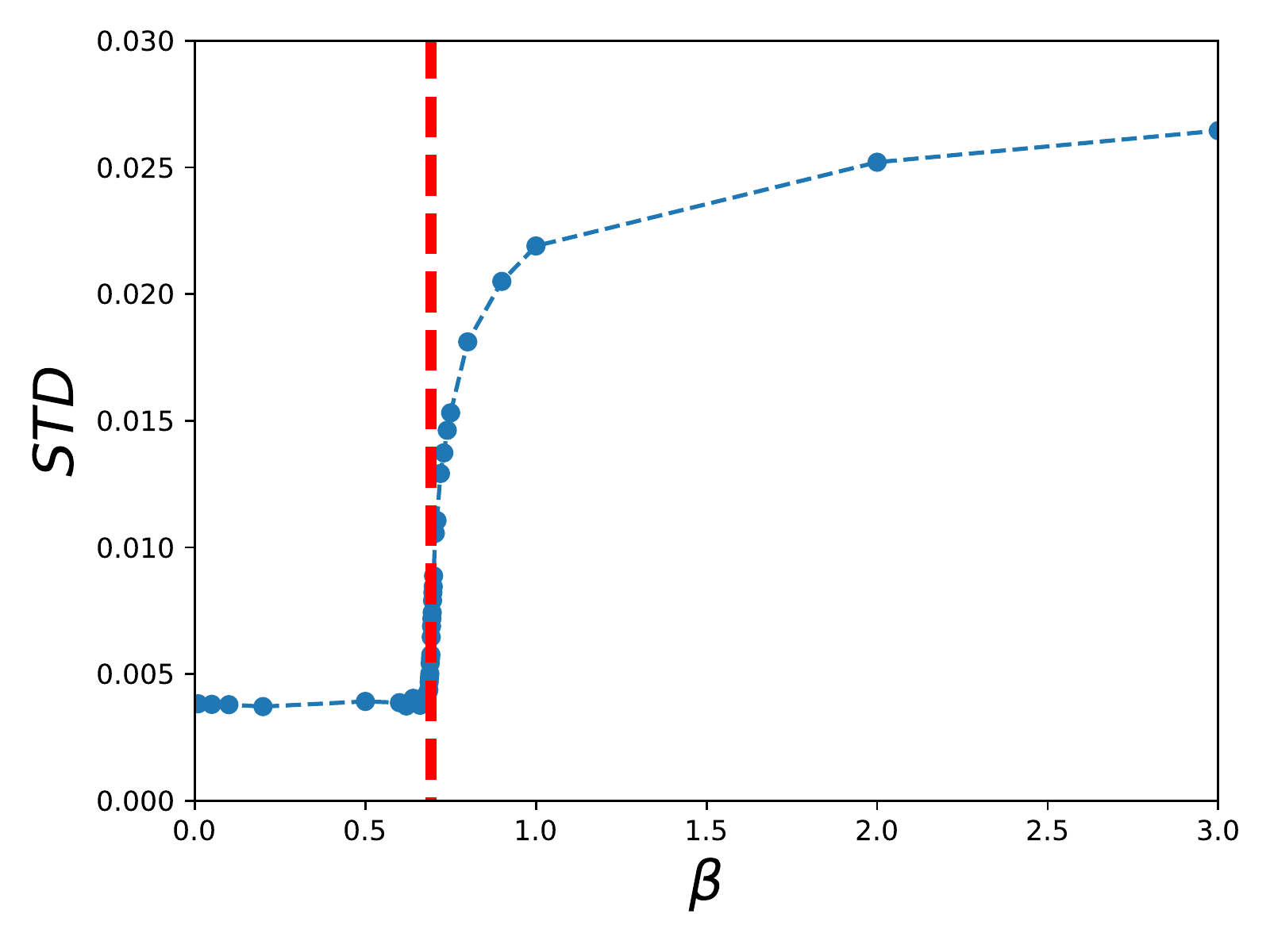}
	\end{center}\vskip-0.7cm
	\caption{The standard deviations (STD) of the GANs outputs as a function of the inverse temperature $\beta$ for the
		3D classical $O(3)$ model. The system size is $L=48$ and the vertical dashed line is the expected $\beta_c$.
		The value of STD begins to rise significantly at $\beta_c$.}
	\label{GANs_O3}
\end{figure}

The standard deviations (STD) of the GANs outputs (The GANs outputs are numbers between 0 and 1) as a function of $\beta$
for the 3D classical $O(3)$ model is shown in fig.~\ref{GANs_O3}.
The linear system size $L$ for the data in the figure is $L=48$.
As expected, the figure demonstrates that the value of STD rises dramatically at $\beta_c$. The shown outcomes indicate the STD of the
NN outputs indeed can be employed to locate the critical point $\beta_c$.  

\subsection{2D generalized $XY$ model}

\subsubsection{AE results}

\begin{figure}
	\begin{center}
		\includegraphics[width=0.5\textwidth]{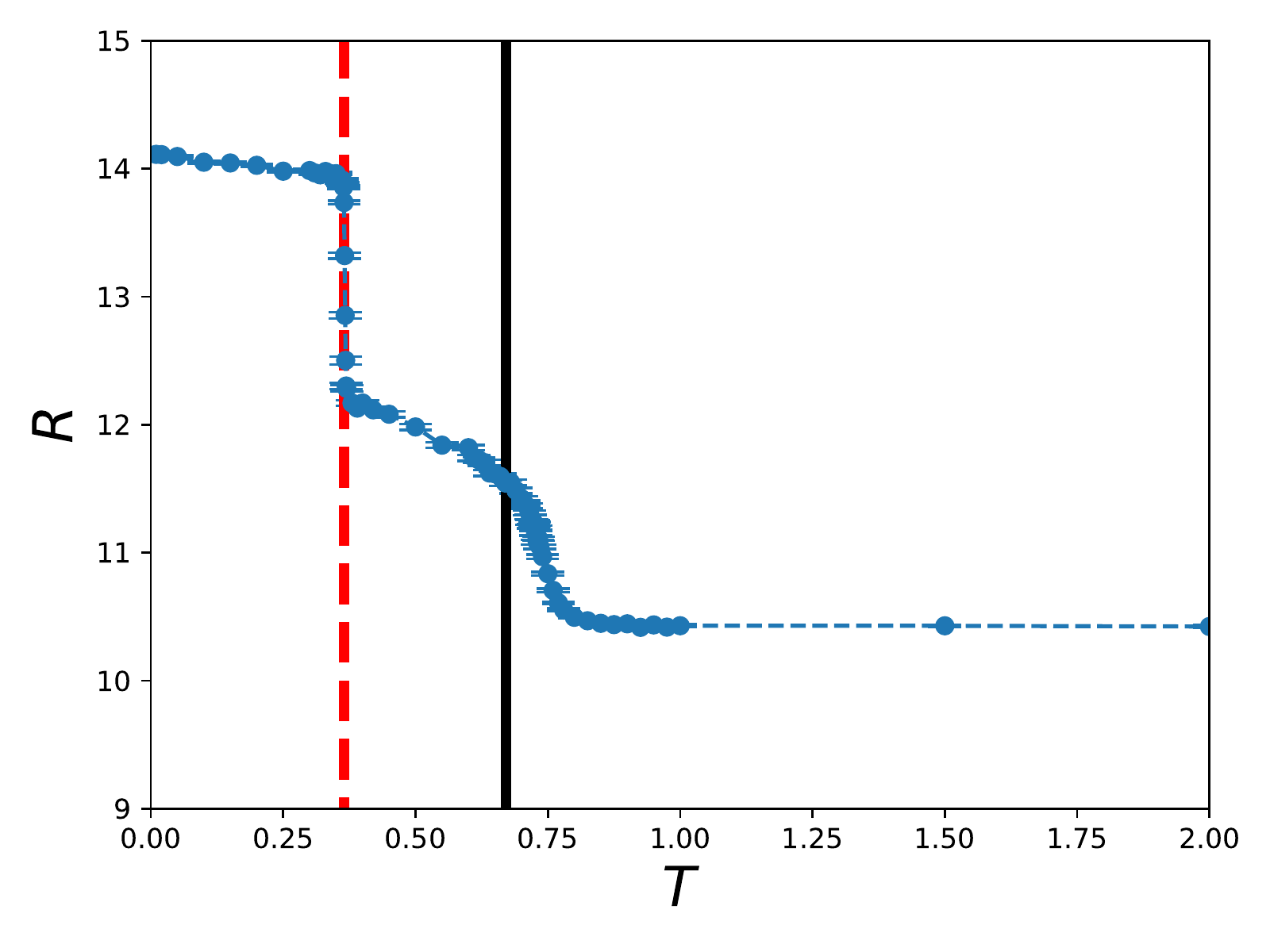}
	\end{center}\vskip-0.7cm
	\caption{The magnitude $R$ of the AE outputs as a function of the temperature $T$ for the
		2D generalized classical $XY$ model. The system size is $L=128$. The vertical dashed and solid lines are the expected $T_c$
		associated with the 3-state Potts and the Berezinskii–Kosterlitz–Thouless (BKT) universalities, respectively.
		The value of $R$ drops obviously at $T_{c,\text{Potts}}$ and $T_{c,\text{BKT}}$.
	}\label{AE_GXY}
\end{figure}

The magnitude $R$ of the AE outputs as a function of the temperature $T$ for the
2D generalized classical $XY$ model is shown in fig.~\ref{AE_GXY}. The system size is $L=128$.
The vertical dashed and solid lines in the figure are the expected $T_{c,\text{Potts}}$ and $T_{c,\text{BKT}}$
associated with the 3-state Potts and the Berezinskii–Kosterlitz–Thouless (BKT) universalities, respectively. As can been seen from
the figure, a sudden jump of $R$ occurs at $T_{c,\text{Potts}}$. Moreover, close $T_{c,\text{BKT}}$ the behavior of
$R$ shows an apparent drop as well. These observed phenomena obviously indicate that the locations of critical points in
the parameter space can be estimated by the $T$-dependence of $R$. It is also interesting to notice that the drop
of $R$ at $T_{c,\text{BKT}}$ is less sharp than that at $T_{c,\text{Potts}}$. This observation is consistent with the fact
that BKT transitions receive certain logarithmic corrections, hence only with extremely large system size will
the signal of the transitions appear transparently.

\subsubsection{GANs results}

\begin{figure}
	\begin{center}
		\includegraphics[width=0.5\textwidth]{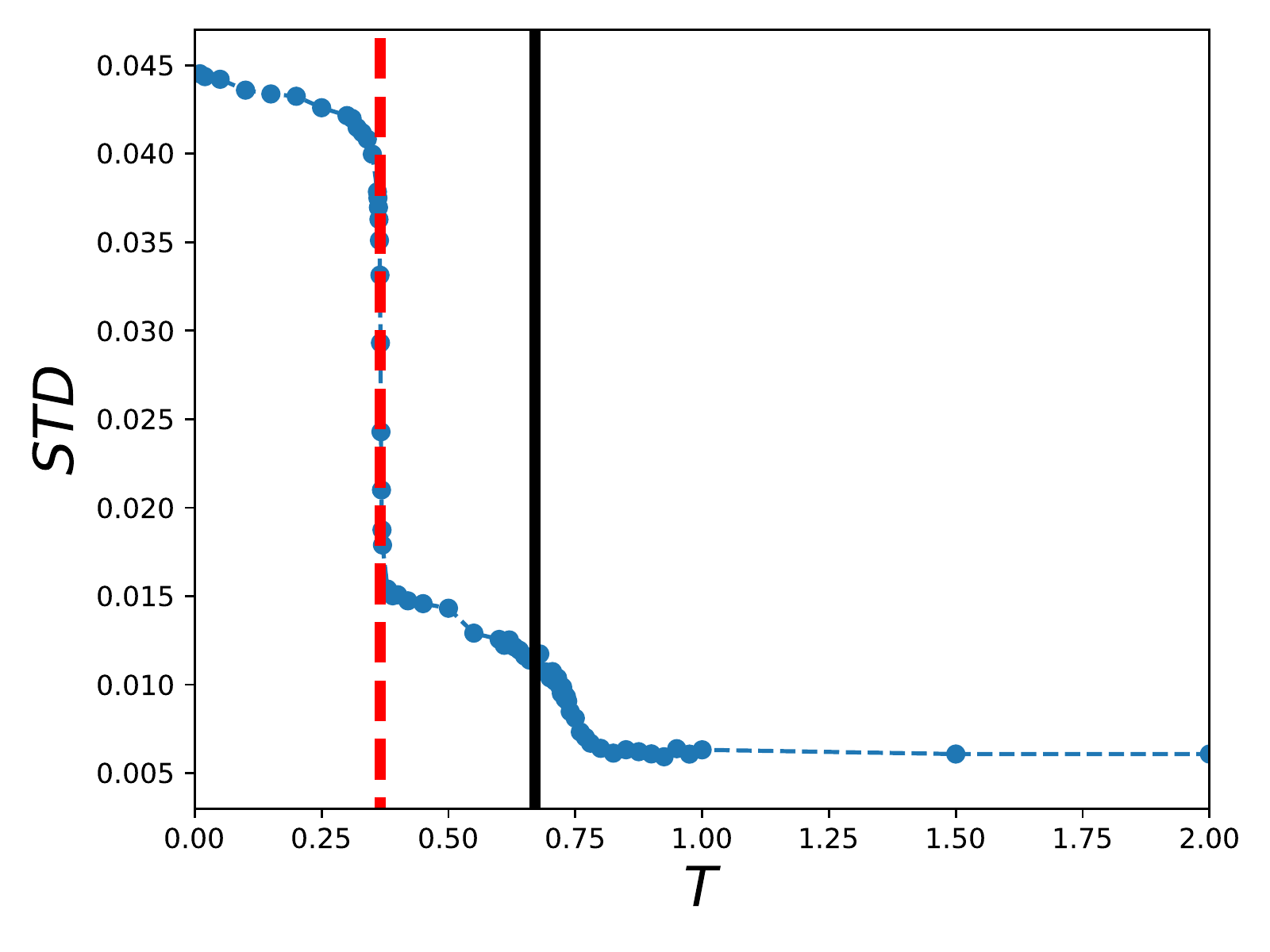}
	\end{center}\vskip-0.7cm
	\caption{The standard deviations of the NN (GANs) outputs as a function of the temperature $T$ for the
		2D generalized classical $XY$ model. The system size is $L$=128. The vertical dashed and solid lines are the expected $T_c$
		associated with the 3-state Potts and the BKT universalities, respectively.
		The value of STD drops obviously at $T_{c,\text{Potts}}$ and
		$T_{c,\text{BKT}}$.}
	\label{GANs_GXY}
\end{figure}

The standard deviations (STD) of the GANs outputs as a function of $T$
for the 2D generalized classical $XY$ model is shown in fig.~\ref{GANs_GXY}.
The linear system size $L$ for the data in the figure is $L=128$.
As expected, the figure demonstrates that the value of STD diminishes dramatically at $T_{c,\text{Potts}}$ and $T_{c,\text{BKT}}$.
We would like to emphasize the fact that when the outcomes in fig.~\ref{GANs_GXY} are compared with that in fig.~\ref{GANs_O3},
the drop of STD at $T_{c,\text{BKT}}$ is less abrupt. This indicates that when BKT transitions are concerned, AE has better
performance of detecting their existence than that of GANs.

\subsection{2D two-state Potts model}

\subsubsection{AE results}

\begin{figure}
	\begin{center}
		\includegraphics[width=0.5\textwidth]{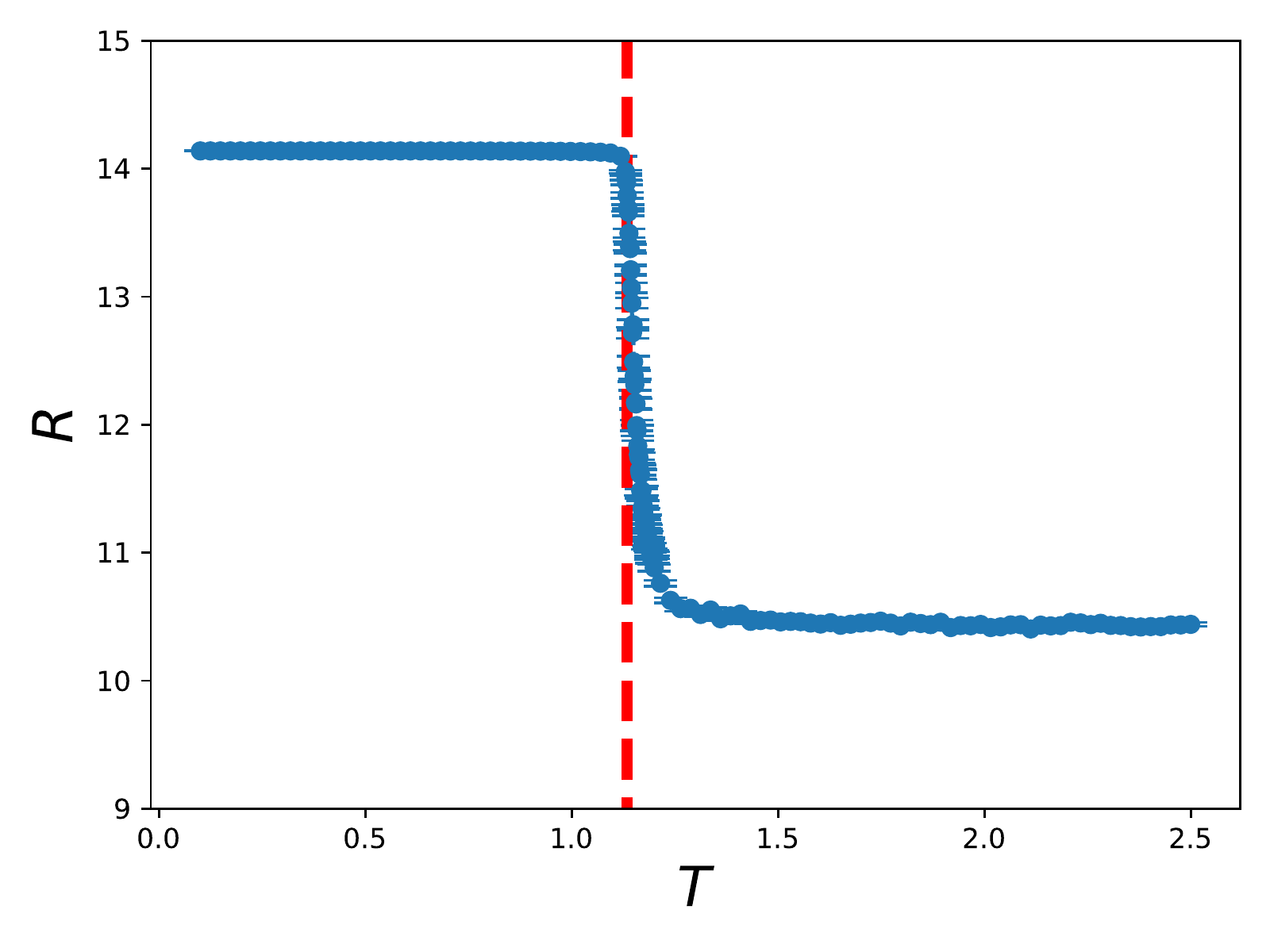}
	\end{center}\vskip-0.7cm
	\caption{The magnitude $R$ of the AE outputs as a function of the temperature $T$ for the
		2D two-state Potts model. The system size is $L$=120. The vertical dashed line is the expected $T_c$.
		The value of $R$ drops dramatically at $T_c$.}
	\label{AE_Potts}
\end{figure}

The magnitude $R$ of the AE outputs as a function of the temperature $T$ for the
2D two-state Potts model is shown in fig.~\ref{AE_Potts}. The system size is $L=120$ and
the vertical dashed line is the expected $T_c$.
Clearly the outcomes demonstrated in the figure imply that $R$ drops significantly at $T_c$.

\subsubsection{GANs results}

\begin{figure}
	\begin{center}
		\includegraphics[width=0.5\textwidth]{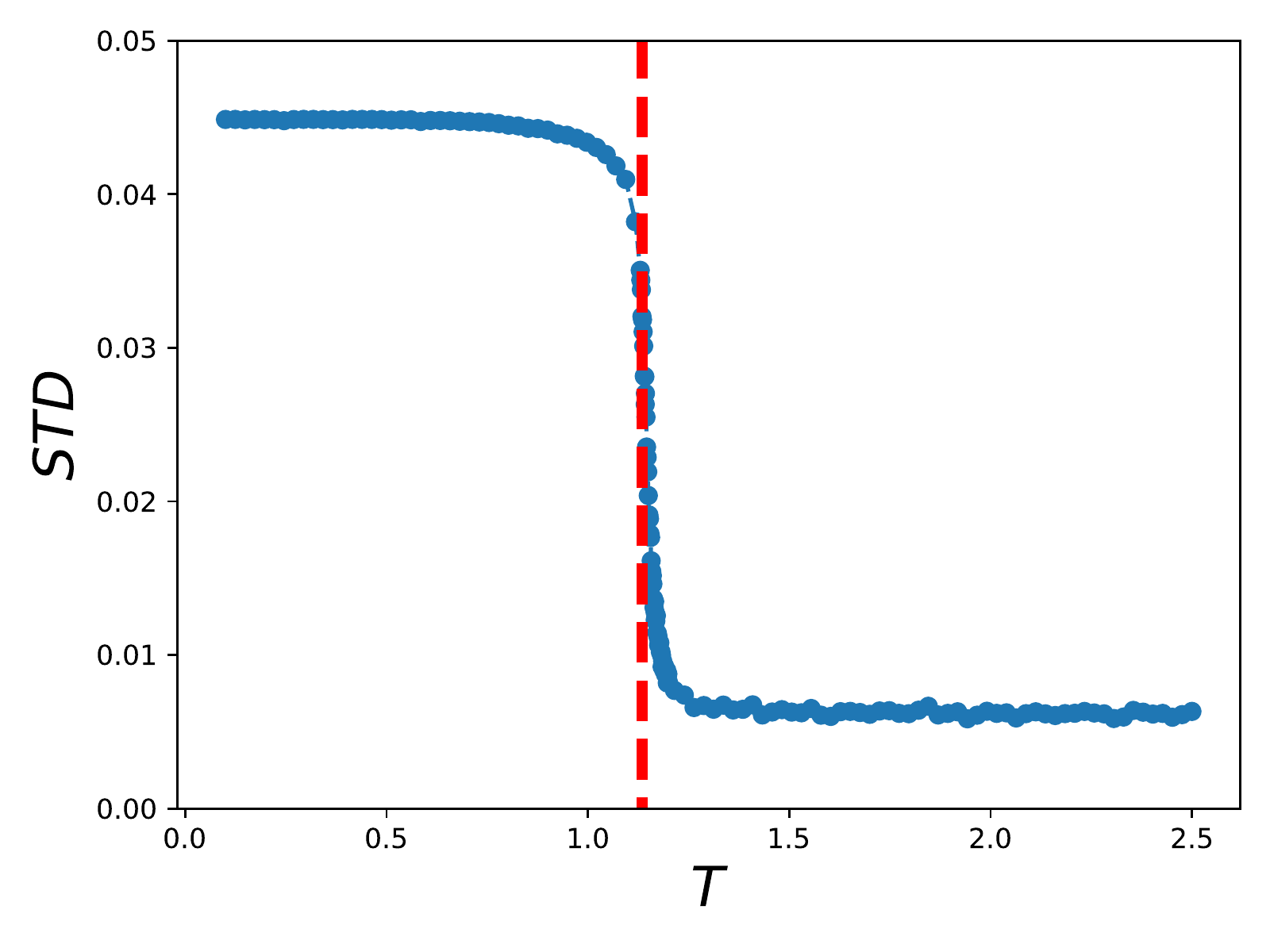}
	\end{center}\vskip-0.7cm
	\caption{The standard deviations STD of the GANs outputs as a function of the temperature $T$ for the
		2D two-state Potts model. The system size is $L=120$ and the vertical dashed line is the expected $T_c$.
		The value of STD drops dramatically at $T_c$.}
	\label{GANs_Potts}
\end{figure}

The standard deviations (STD) of the GANs outputs as a function of $T$ for the two-state Potts model
is shown in fig.~\ref{GANs_Potts}.
The linear system size $L$ for the data in the figure is $L=120$ and the vertical dashed line is the expected $T_c$.
As expected, the figure demonstrates that the value of STD drops dramatically at $T_c$.

\subsection{1D Bose-Hubbard model}

The associated simulations are conducted with system size $L=256$, $t=0.1$, $U=1.0$, $T=0.0025$, and various values of $\mu$. 
As $\mu$ varies, one expects to see a transition from the superfluidity to the Mott insulator. In addition, the maximum number of Bose 
per site is set to 5, and the naive configurations used for the NN
prediction are built from the local densities which are generated based on the means and the mean errors of the outcomes from 
the Monte Carlo simulations. 
Finally, in the process of one-hot encoding, if the local density $n_i$ of an original site $i$ is greater (smaller) than 0.99 (i.e. 
we set a density cut off to be 0.01), then integer 1 (0) is assigned to the associated spot.

\subsubsection{AE results}

\begin{figure}
	\begin{center}
		\includegraphics[width=0.5\textwidth]{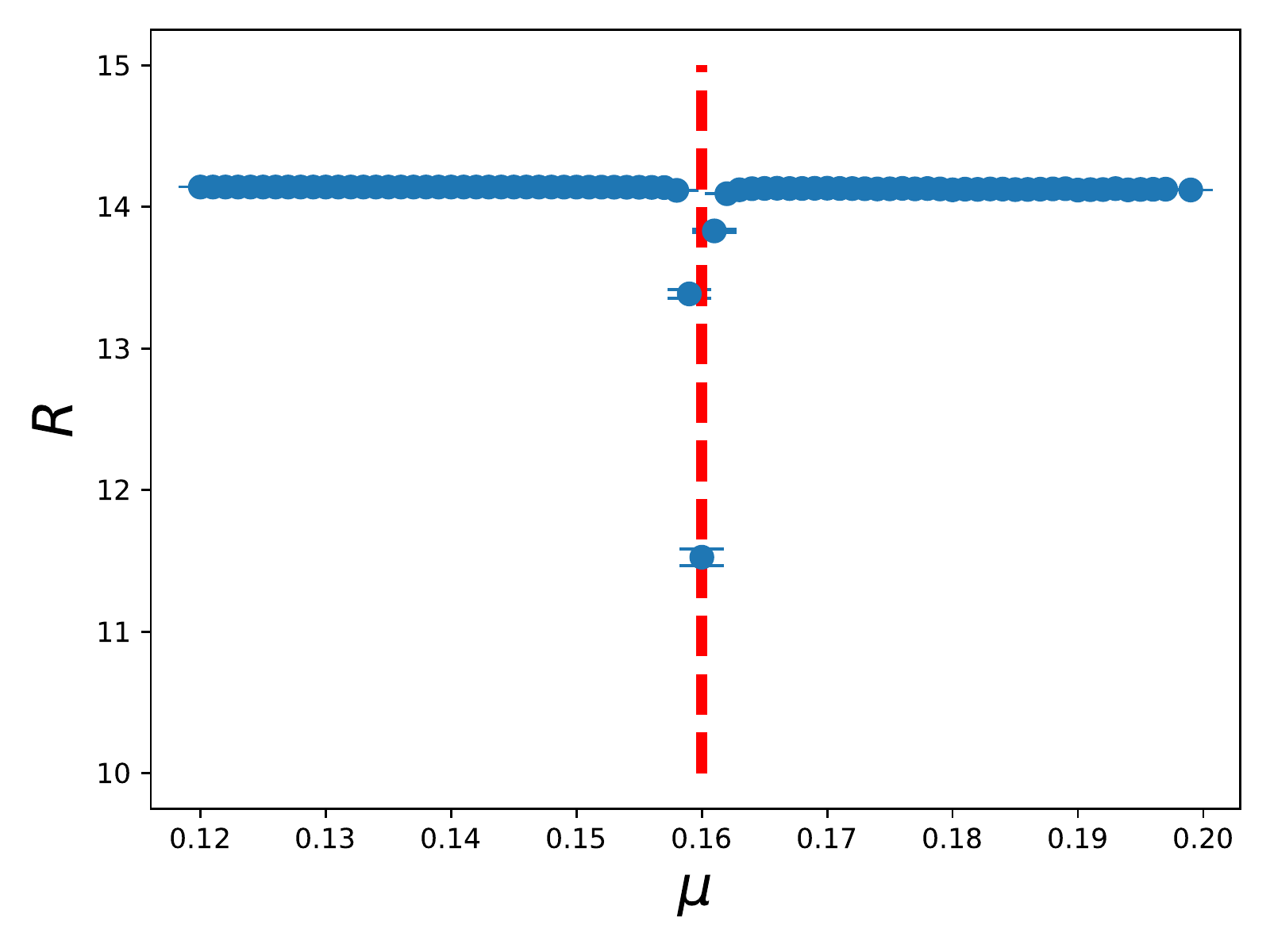}
	\end{center}\vskip-0.7cm
	\caption{The magnitude $R$ of the AE outputs as a function of $\mu$ for the 1D Bose-Hubbard model. The vertical dashed line is the expected
	$\mu_c$.}
	\label{AE_HB}
\end{figure}

The magnitude $R$ of the AE outputs as a function of $\mu$
is shown in fig.~\ref{AE_HB}. The dashed vertical line in the figure is the critical point $\mu_c$ 
estimated from Refs.~\cite{Eji11,Eji12}.
As can been seen from the figure, the $\mu$ which has the smallest value of $R$ is very close to $\mu_c$. In other words,
the constructed AE is capable of detecting the considered phase transition of the 1D Bose-Hubbard model. Here we would like
to point out that if one uses another value of density cut off, then the $\mu$ having the minimum $R$
clearly will shift (slightly). This will have certain impact on estimating $\mu_c$. It is obviously that such an impact will
become much milder if extremely accurate data points (so that the density cut off can be very tiny) are used for the
(NN) calculations.

\subsubsection{GANs results}

\begin{figure}
	\begin{center}
		\includegraphics[width=0.5\textwidth]{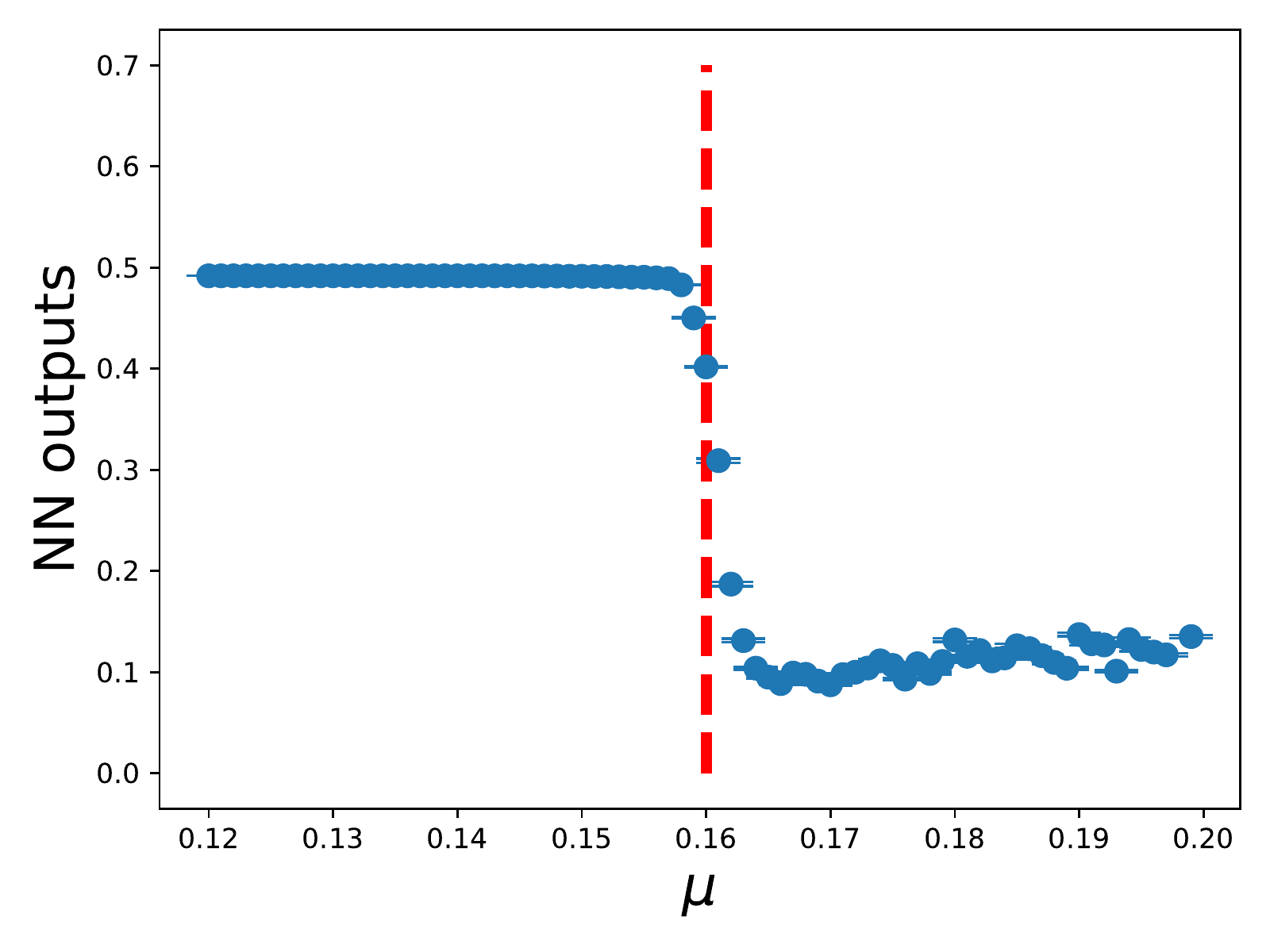}
	\end{center}\vskip-0.7cm
	\caption{The GANs outputs as a function of $\mu$ for the 1D Bose-Hubbard model. The vertical dashed line is the expected $\mu_c$.}
	\label{GANs_BH}
\end{figure}

For the GANs, when the detection of the phase transition (of the 1D Bose-Hubbard model) is concerned, 
the performance of the outputs themselves is better than that of the standard deviation. Indeed,
as can been seen from fig.~\ref{GANs_BH}, the $R$ drops significantly when one approaches $\mu_c$ (the vertical dashed line) from the left. 
Here again a density cut off of 0.01 is considered. It should be pointed out that unlike those used in the previous sections, the GANs leading to
the results present in fig.~\ref{GANs_BH} has been trained with the additional step of one-hot encoding.

\section{Discussions and conclusions}

In this study, an AE having only one hidden layer of two neurons as well as
a GANs with 2 neurons generator and 2 neurons discriminator 
are constructed. In particular, they are trained only once on a 1D lattice of 200 sites.
The training set used consists of two artificial configurations. The obtained unsupervised NNs successfully determine the critical points
of several models, including the 3D classical $O(3)$ model, the 2D generalized
classical $XY$ model, the 2D two-state Potts model, as well as the 1D
Bose-Hubbard model. It is remarkable that an AE and a GANs with such simple
architectures can lead to high precision calculations of the targeted phase
transitions. It is beyond doubt that with moderate modifications, the
applications of the built AE and GANs here
can be extended to include the Fermi-Hubbard or SSH models. It is
amazing as well that any of the two unsupervised NNs built here 
can be employed to calculate the critical points of various 3D and 2D models
accurately. In particular, the constructed AE and GANs are able to
successfully detect both the symmetry breaking and the topology related phase
transitions. It is also surprising that no information of these models
is needed for the built AE and GANs to detect the associated phase
transitions. 

When compared with the conventional deep learning AEs typically used in the
literature, a benchmark investigation indicates that at least several thousand
speed up is gained for the AE employed here. Certain
gain in the calculations of prediction conducted here is
anticipated as well. Finally, we would like to mention that the training of the GANs
considered in this study is also extremely efficient. Specifically, it takes about
154 seconds to train the GANs (with 2000 epochs) \cite{time1}.

It should be pointed out that in Ref.~\cite{Kim18}, a supervised NN with one
hidden layer of two neurons has successfully determined the $T_c$ of the 2D Ising
model. Here we have gone much further than that achieved in Ref.~\cite{Kim18}.

For each of the models studied here, only one system size data is considered.
Definitely one can carry out the needed calculations to obtain results of several
box sizes. With these new outcomes, semi-experimental finite-size scaling ansatz can be
utilized to reach an estimation of the bulk critical point(s) \cite{Tan20.2,Tse22}. Here we do not
perfrom such an investigation since relevant detailed studies are available in
Refs.~\cite{Tan20.2,Tse22}.

We would also like to emphasize the fact that the conventional training methods that usually
used in the literature is computational demanding and require huge amount of computer memory. Hence,
most of the calculations are limited to small to moderate large system sizes. Similar to the supervised NN
considered in Refs.~\cite{Tan20.2,Tse22}, there is no system size restriction
for the built AE and GANs employed here.

In addition to the advantage of extremely efficient training, with the method adopted here much less storage capacity is required
when compared with the standard NN approaches used in the literature as well. Practically one only needs to save 200 spins for every
generated large system size configuration which usually comes with several thousand to over 10 thousand spins.
This feature is similar to the supervised NN employed in Ref.~\cite{Tan20.2,Tse22}. 
To demonstrate the high efficiency of
the employed unsupervised NNs, we have recorded the total time needed to complete the NN calculations associated with the 3D
$O(3)$ model. The 3D $O(3)$ model is chosen because it takes the longest time to perform the related NN investigation 
for this model using the conventional approaches. For the AE considered in this study,  
from the start of the training to the end of the prediction the time needed for these procedures is about 200 seconds (This also
includes the time required for reading (loading) the data files). In other words,
it takes less than 4 minutes to produce the results shown in fig.~\ref{AE_O3}. Here for each of the 49 temperatures, 
two thousand configurations are
used for the NN prediction. In reality, one can only keep two hundred relevant quantities (such as the spins) for 
every generated configurations and uses these data to perform the NN prediction with the AE and GANs constructed here. 
As a result, it is anticipated that if the same amount of data are employed, 
then the required time to 
complete the whole NN calculations for each of the other studied models is about the same as that of the 3D $O(3)$ model.
 
Based on the results obtained here as well as that in Ref.~\cite{Tan21}, it is likely
that the training schemes introduced here and in Ref.~\cite{Tan21} are the most 
efficient strategy for training when studying phase transitions with NN are
concerned. In addition, these outcomes show that a NN trained once with
two artificial configurations can be applied to many models that are 
different from each other dramatically. This suggests strongly that
when machine learning is considered, many, or even the majority of phase
transitions belong to a class having two elements, namely the Ising class.

Finally, it will be interesting to examine whether by simply engineering the configurations
for prediction one can apply the unsupervised NN constructed here to study the criticalities of some other models that are
not investigated in this study.
It is also interesting to see if similar elegant ideas exist for
other research fields of physics.

\section*{Acknowledgements}


\paragraph{Funding information}
Partial support from the Ministry of Science and Technology of Taiwan (MOST) is acknowledged
(MOST 110-2112-M-003-015 and MOST 108-2112-M-029-006-MY3).




\begin{thebibliography}{99}
\bibitem{time1}
	It takes about 388 seconds to train the GANs with the step of one-hot encoding.
\end{thebibliography}


\end{document}